\documentclass[a4paper, 11pt]{article}

\usepackage{fullpage}

\usepackage{graphicx}
\usepackage{subcaption}

\usepackage{amssymb}

\usepackage{lineno}
\usepackage{amsmath,amsthm,amsfonts}
\usepackage{url}
\usepackage{tikz}
\usepackage{booktabs}

\usepackage[numbers,sort&compress]{natbib}

\usepackage{framed} 
\usepackage{multicol} 
\usepackage{multirow}
\usepackage{nomencl} 
\usepackage{authblk}

\newcommand\nc{\newcommand}
\nc\pa{\partial}
\nc\pad[2]{\frac{\pa #1}{\pa #2}} 
\nc\padd[2]{\frac{\pa^2 #1}{\pa
{#2}^2}} 
\nc\nd[2]{\frac{\text{d} #1}{\text{d} #2}}
\nc\ndd[2]{\frac{d^2 #1}{d {#2}^2}}
\nc\pat[2]{\frac{D #1}{D
#2}} 
\nc{\q}{\mathbf{Q}}
\nc{\ii}{\text{\textbf{i}}}
\nc{\Kn}{\text{Kn}}
\nc\ra{\rightarrow} \nc\Ra{\Rightarrow} 
\nc{\ud}{\mathrm{d}}

\title{A slip-based model for the size-dependent effective thermal conductivity of nanowires}

\author[1,2]{M. Calvo-Schwarzw\"alder}
\author[1]{M.~G. Hennessy}
\author[3]{P. Torres}
\author[1,2]{T.~G. Myers}
\author[3]{F.~X. Alvarez}

\affil[1]{Centre de Recerca Matem\`atica, Barcelona, Spain.}
\affil[2]{Departament de Matem\`atiques, Universitat Polit\`ecnica de Catalunya, Barcelona, Spain.}
\affil[3]{Departament de F\'isica, Universitat Aut\`onoma de Barcelona, Barcelona, Spain.}

\providecommand{\keywords}[1]{\small\textbf{\textit{Keywords:}} #1}

\begin{document}
 
\maketitle

\begin{abstract}
The heat flux across a nanowire is computed based on the
Guyer-Krumhansl equation. Slip conditions with a slip length depending
on both temperature and nanowire radius are introduced at the outer boundary. An
explicit expression for the effective thermal conductivity is derived
and compared to existing models across a given temperature range,
providing excellent agreement with experimental data for Si nanowires.
\\ \newline
\noindent\keywords{Heat Transfer, Nanotechnology, Thermal Conductivity, Guyer-Krumhansl, Phonon Hydrodynamics}

\end{abstract}



\section{Introduction}
\label{S:1}
Nanotechnology is currently the focus of extensive research due to its wide range of applications in
fields such as industry and medicine
\cite{Salata2004,Ahmad2012,Cregan2015,Garnett2011,Liu2015,Ge2012,Chinen2015}. Nanowires, in
particular, are being used in technologies relating to solar cells \cite{Garnett2011}, flexible
screens \cite{Liu2015}, detection of cancerous cells \cite{Chinen2015}, and energy storage
\cite{Ge2012}.  A key issue facing the practical use of nanodevices is thermal management
\cite{Cahill2014}.  Inefficient regulation of heat can lead to large temperatures and melting,
possibly resulting in device failure.  Understanding and predicting heat flow on the nanoscale is
therefore crucial for the manufacturing and operation of nanotechnologies.

It is widely known that many thermophysical material properties become size-dependent at the
nanoscale \cite{Buffat1976,Sun2007,Tolman1949,Xiong2011,Lai1996,Li2003,Wronski1967,Shin2014}. Buffat
and Borel \cite{Buffat1976} showed a dramatic decrease of the melting temperature of gold
nanoparticles of almost 50\% from the bulk value. For aluminium nanoparticles, a decrease in latent
heat by a factor of four has been reported \cite{Sun2007}. Experimental observations also
demonstrate that the thermal conductivity in silicon nanowires is much lower than the theoretical
value predicted by kinetic theory \cite{Li2003}. For instance, it is reported that, at room
temperature, the thermal conductivity of Si nanowires with a diameter of 37 nm decreases by
approximately 87\% with respect to the bulk value. Furthermore, when the characteristic size of the
system is much smaller than the phonon mean free path, the thermal conductivity shows an
approximately linear dependence on size \cite{Ma2012,Alvarez2008,Tzou2011,Alvarez2009}.

The size dependence of the thermal conductivity of nanosystems is attributed to the fact that, on
the nanoscale, the transport of thermal energy is a ballistic process driven by infrequent
collisions between thermal energy carriers known as phonons. This is in contrast to macroscopic heat
transfer, which is a diffusive processes driven by frequent phonon collisions. As the size of a
device becomes commensurate with the phonon mean free path, bulk phonons are more likely to collide
with a boundary than with each other. The ability of a nanodevice to conduct thermal energy,
therefore, becomes strongly influenced by the scattering dynamics at the boundary as well as the
geometrical structure (e.g., size and shape) of this boundary.

Due to the fundamentally different manner in which heat is transported across nanometer length
scales in comparison to heat flow at the macroscale, Fourier's law is unable to provide an accurate
description of heat conduction in this regime \cite{Chang2008}. Different approaches to modelling
nanoscale heat flow have been developed in order to capture the ballistic nature of energy transport
and size dependence of the effective thermal conductivity (ETC). These approaches can be classified
into three main categories: microscopic, mesoscopic and macroscopic models. Microscopic approaches,
such as molecular dynamics or Monte-Carlo methods \cite{Chen2005}, focus on the evolution of every
single phonon while mesoscopic models group them together depending on their wavelength and
wavevector. Micro and mesoscopic models are mainly based on the Boltzmann transport equation (BTE)
and its solution under different approximations. A popular example is the equation of phonon
radiative transfer (EPRT) \cite{Majumdar1993}, where an expression for the ETC similar to the
classical expression from kinetic theory is derived, although here an effective mean free path is
now considered. However this model is based on the gray approximation and thus considers a single
phonon group velocity and lifetime. A more general model where these quantities are mode-dependent
was presented by McGaughey \emph{et al.}~\cite{McGaughey2011}. Starting again from the BTE, Alvarez
\emph{et al.}~\cite{Alvarez2008,Alvarez2007} extract a continued-fraction expression to describe the
ETC in thin films. Other models, such as those of Callaway \cite{Callaway1959} and Holland
\cite{Holland1963}, also consider phonon distributions rather than single phonons, but Mingo
\emph{et al.}~\cite{Mingo2003} showed that they fail when predicting the ETC for Si or Ge
nanowires.

Macroscopic models aim to describe global variables of the system, such as the temperature and the
heat flux. A recent approach at the macroscopic scale is the thermomass model, where heat carriers
are assumed to have a finite mass determined by Einstein's mass-energy relation
\cite{Tzou2011,Wang2010,Wang2014}. Other approaches are based on the Guyer-Krumhansl (G-K) equation
\cite{Guyer1966a,Guyer1966b}, which is derived from a linearized BTE in dielectric crystals. This
equation has become popular since it is analogous to the extensively studied Navier-Stokes (N-S)
equations and it is one of the simplest extensions to Fourier's law that includes memory and
non-local effects. Models based on the G-K equation are included in the framework of phonon
hydrodynamics \cite{Jou1996}. For instance, Alvarez \emph{et al.}~\cite{Alvarez2009} use the analogy
between the G-K equation and the N-S equations to derive an expression for the ETC in circular
nanowires, splitting the heat flux into two separate contributions. This work has been extended to
elliptical and rectangular nanowires \cite{Sellito2012}. However, since the size of the phonon mean
free path depends on temperature, the assumptions on which they base their reductions are only valid
for low temperatures or very small sizes. Dong \emph{et al.}~\cite{Dong2014} find an expression for
the ETC by solving the full G-K equations at steady state, although their no-slip boundary
condition leads to a quadratic dependence of the ETC on the characteristic size of the device for
large Knudsen numbers instead of the known linear behaviour. This does not happen in the case of the
expression derived by Ma \cite{Ma2012}, where a fixed flux is imposed on the outer
boundary. However, Ma's solution for the nanowire shows a very poor match to data (their paper
contains an error in that they plot the thin film solution in their figure for the nanowire; this
actually shows reasonable agreement with the data). Further, most of the existing models are only
validated at room temperature and therefore a deeper assessment of their accuracy is required.

In this paper, we introduce a new phenomenological slip boundary condition that, when used with the
G-K equations, results in ETC predictions that are in excellent agreement with experimental data for
Si nanowires over a range of radii and temperatures. The proposed model is remarkably simple and
only requires knowledge of the temperature dependence of the bulk thermal conductivity and phonon
mean free path, both of which may be obtained experimentally or computationally, making it well
suited for use in practical applications. A detailed comparison of the proposed and existing models
is performed, the results of which show that the proposed model consistently yields the most
accurate predictions of the ETC compared to existing models. 
Furthermore, this comparison establishes the validity of each model
in terms of temperature and nanowire radius.

\section{Mathematical modelling}

\begin{figure}
\centering
\begin{tikzpicture}
\fill [left color=red!80, right color=red!50,opacity=0.3] (0,0) ellipse (.5 and 1.25);
\fill [left color=blue!20, right color=blue!50,opacity=0.3] (5,0) ellipse (.5 and 1.25);
\draw (0,0) ellipse (.5 and 1.25);
\draw (0,1.25) -- (5,1.25);
\draw [dashed] (5,-1.25) arc (270:90:0.5 and 1.25);
\draw (5,1.25) arc (90:270:-.5 and 1.25);
\draw (0,-1.25) -- (5,-1.25);
\fill [left color=red!50, right color=blue!50,opacity=0.3] (0,-1.25) -- (5,-1.25) arc (270:90:-.5 and 1.25) -- (0,1.25) arc (270:90:-.5 and -1.25);
\draw [<->] (0,0) -- (0,1.2);
\node at (-.2,.6) {$R^*$};
\draw [<->] (0, -1.5) -- (5, -1.5) node [midway, below,fill=white] {$L^*$};
\draw [->] (2,0) -- (3,0) node [midway, below] {$x^*$};
\draw [->] (2,0) -- (2,1) node [midway, left] {$r^*$};
\draw [->] (-2,.5) -- (-1,.5) node [midway, above] {$\q^*$};
\draw [->] (-2,0) -- (-1,0);
\draw [->] (-2,-.5) -- (-1,-.5);
\node at (0,1.5) {$T^*=T^*_0$};
\node at (5,1.5) {$T^*=T^*_1$};
\end{tikzpicture}
\caption{A circular nanowire with radius $R^*$ and length $L^*$ is held at different temperatures $T_0^*$, $T_1^*$ at the left and right ends respectively, which induces a heat flux $\q^*$.}
\label{fig:cylinder}
\end{figure}
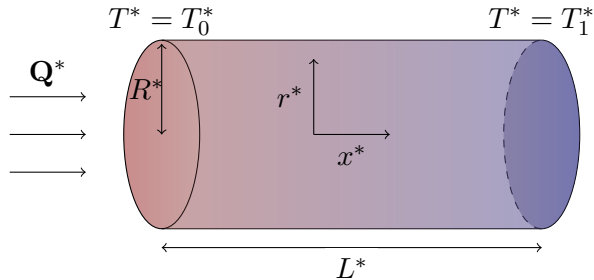

We consider a circular nanowire (NW) of radius $R^*$ and length $L^*$ that is suspended in a vacuum;
see Fig.~\ref{fig:cylinder}. The radius of the NW is assumed to be much smaller than its length,
i.e., $R^* / L^* \equiv \epsilon \ll 1$. A temperature gradient $\Delta T = T_0^* - T_1^* > 0$ is
imposed along the axial direction of the NW by fixing the temperature at its left and right ends to
be $T_0$ and $T_1$, respectively. The thermal flux that is driven by this temperature gradient is
assumed to be axisymmetric. Therefore, it is sufficient to consider a two-dimensional model with
radial and axial coordinates $r^*$ and $x^*$, respectively.  The mathematical model will consist of
an equation representing conservation of thermal energy and the G-K equation describing the
evolution of the thermal flux.

\subsection{Bulk equations}

Conservation of energy requires 
\begin{equation}\label{cons_energy}
	\pad{\mathfrak{u}^*}{t}=-\nabla\cdot\q^*,
\end{equation}
where $\mathfrak{u}^*(T^*)$ is the internal energy per unit mass and $T^*$ is the temperature.  The
thermal flux $\q^* = v^*\mathbf{\hat r}+w^*\mathbf{\hat x}$ is assumed to satisfy the G-K equation
\begin{equation}\label{gk}
	\tau^*\pad{\q^*}{t}+\q^*=-k^*\nabla T^*+\ell^{*2}\left(\nabla^2\q^*+2\nabla\nabla\cdot\q^*\right),
\end{equation}
where $v^*$ and $w^*$ are the radial and axial components of the heat flux, $\tau^*(T^*)$ is the
total mean free time, $k^*(T^*)$ is the bulk thermal conductivity in the Kinetic Collective Model
(KCM) framework \cite{Torres2016}, and $\ell^*(T^*)$ is a non-local length related to the bulk
phonon mean free path (MFP), i.e., the mean distance between phonon-phonon collisions. Zhu \emph{et
al.}~\cite{Zhu2017} propose that $\ell^*$ is the geometric mean of the bulk MFP and a local MFP, the
latter of which decreases near a boundary. Here, we opt for simplicity and model the decrease in MFP
near the boundary through the slip boundary condition.  For convenience, we will not write the
temperature dependence of the parameters explicitly unless it is required due to the
context. However, we note that, for the temperature ranges considered here, both the bulk thermal
conductivity and non-local length monotonically decrease with temperature. As will be shown in
Sec.~\ref{sec:valid}, the non-local length of silicon decreases from about 5 $\mu$m at 50 K to 55 nm
at 300 K.

The first term on the left-hand side of \eqref{gk} captures memory effects and, in particular, the
dependence of heat flux on the history of the temperature gradient. The second term on the
right-hand side of \eqref{gk} captures non-local effects, such as the interaction of phonons with
the boundary of the NW. When the characteristic time and length scales are much larger than the
resistive mean free time and non-local length, the G-K equation \eqref{gk} reduces to Fourier's law.

For the remainder of the paper, we restrict our attention to the case of steady-state heat
flow. This focus is motivated by the available experimental data. Under the steady-state assumption,
$\partial \mathfrak{u}^*/\partial t =\partial\q^*/\partial t =0$. Conservation of energy
\eqref{cons_energy} and the G-K equation \eqref{gk} then reduce to
\begin{subequations}
  \label{gk_steady}
  \begin{align}
    \nabla\cdot\q^*&=0, \label{dim:steady_energy}\\
    \q^*&=-k^*\nabla T^*+\ell^{*2}\nabla^2\q^*. \label{dim:steady_gk}
  \end{align}
\end{subequations}
This system is clearly analogous to an incompressible, viscous flow with a source term proportional
to the velocity. Under this analogy, the parameter $\mu^*=\ell^{*2}/k^*$ may be interpreted as a thermal
viscosity. These observations allow us to use well-known techniques from viscous flow to analyse the
problem.

\subsection{Boundary conditions}\label{sec:bc}

The boundary conditions for the temperature at the endpoints of the NW are
\begin{equation}
	\left.T^*\right|_{x^*=0}=T_0^*, \qquad \left.T^*\right|_{x^*=L^*}=T_1^*,
\end{equation}
where $T_0^*>T_1^*$. Due to the symmetry of the problem the boundary conditions at $r^*=0$ are
straightforward,
\begin{equation}\label{bc_r0}
	\left.v^*\right|_{r^*=0}=\left.\pad{w^*}{r^*}\right|_{r^*=0}=0.
\end{equation}
We assume that no heat flows across the outer boundary, i.e.,
\begin{equation}
	\left.v^*\right|_{r^*=R^*}=0.
\end{equation}
Finally, on the edge of the NW we continue the analogy with fluid dynamics and employ a slip condition,
\begin{equation}
	\left.w^*\right|_{r^*=R^*}=-\ell^*_s\left.\pad{w^*}{r^*}\right|_{r^*=R^*},
        \label{bc:dim_slip}
\end{equation}
where $\ell_s^*$ is the slip length.  In accordance with previous authors (see, e.g.,
Refs.~\cite{Alvarez2009,Zhu2017}), we will assume that the slip length is proportional to the
non-local length, i.e., $\ell_s^*=C\ell^*$, where $C$ is a dimensionless parameter that encodes
detailed information about phonon-boundary scattering and surface roughness.  This form of boundary
condition is discussed in more detail in Refs.~\cite{Alvarez2009,Sellito2010}. The basic idea is
that a slip condition can capture crucial contributions to the thermal flux from reflected phonons.
A no-slip condition neglects such contributions, resulting in predictions of the thermal flux and
hence ETC that decrease too rapidly with the radius of the NW.

Various forms for the parameter $C$ appear in the literature.  Alvarez \emph{et
al.}~\cite{Alvarez2009} write $C$ in terms of the specular parameter $p$, which describes the
precise nature of phonon scattering (i.e., specular or diffuse).  A similar approach is presented in
Zhu \emph{et al.}~\cite{Zhu2017}, although the authors essentially treat $p$, and hence $C$, as a
fitting parameter.  Sellitto \emph{et al.}~\cite{Sellito2010} write $C$ in terms of the surface
geometry and then expand it as a power series in the temperature, resulting in a model with an
excessive number of fitting parameters.  Here, we take a more practical approach and write $C$ as a
simple exponential, $C = \exp(-R^* / \ell^*$), which is similar in form to the expression derived
for and used in the specific case of a two-dimensional rectangular nanolayer by Zhu \emph{et
al.}~\cite{Zhu2017}.  The proposed exponential form does not involve any fitting parameters and it
inherits a temperature dependence through the bulk non-local length $\ell^*(T^*)$.  The motivation
for this expression is as follows. If the non-local length is much less than the NW radius, $\ell^*
\ll R^*$, then phonons are more likely to collide with each other than with the boundary and the
influence of reflected phonons will be small, corresponding to a no-slip condition.  Conversely, if
the non-local length is much larger than the NW radius, $\ell^* \gg R^*$, then phonon-boundary
scattering is more likely than phonon-phonon collisions.  In this case, phonons reflections will
strongly influence the flow, which is captured in a slip condition.

\subsection{Reduction of the governing equations}
We may reduce the governing equations by exploiting the separation between the axial and radial
length scales. First, we introduce non-dimensional variables
\begin{equation}\label{var_ND}
  r = \frac{r^*}{R^*}, \qquad x = \frac{x^*}{L^*}, \qquad  v = \frac{v^*}{v_0^*}, \qquad  w = \frac{w^*}{w_0^*},  \qquad
 T = \frac{T^*-T_1^*}{\Delta T},\qquad k=\frac{k^*}{k_0^*},
\end{equation}
where $v_0^*$ and $w_0^*$ are (unknown) typical values of the axial and radial components of the
heat flux and $k_0^*$ is a reference value for the bulk thermal conductivity. A relationship between
$v_0^*$ and $w_0^*$ is obtained by requiring both terms in \eqref{dim:steady_energy} to have the
same magnitude, which ensures that energy is conserved in the leading-order problem. This yields
$v_0^* =\epsilon w_0^*$. Finally, the classical scale for the axial flux is chosen, $w_0^* = k_0^*
\Delta T / L^*$, which ensures that Fourier's law is recovered in the classical limit, i.e., when
the Knudsen number $\Kn = \ell^* / R^*$ tends to zero.

After non-dimensionalising the model given by \eqref{gk_steady} and neglecting small terms of size
$O(\epsilon)$ and smaller, the governing equations reduce to
\begin{subequations}
\begin{align}
 	&\pad{T}{r}= 0,\label{eq_v_ND} \\
	&w=-k\pad{T}{x}+\frac{\Kn^2}{r}\pad{}{r}\left(r\pad{w}{r}\right), \label{eq_w_ND} \\
  	&\frac{1}{r}\pad{}{r}(rv)+\pad{w}{x}=0.
\end{align}
\end{subequations}
Equation \eqref{eq_v_ND} indicates $T\approx T(x)$ (to the order of the neglected terms), which
simplifies the integration of Eqn.~\eqref{eq_w_ND}. Notice also that if $\Kn\ll 1$ then the
classical (non-dimensional) Fourier's law is retrieved,
\begin{equation}\label{eq_w_Knsmall}
	w_\text{F}=- k\nd{T}{x}.
\end{equation}
The boundary conditions become
\begin{subequations}
\begin{align}
	&\left.v\right|_{r=0}=0, \qquad\left.v\right|_{r=1}=0,\label{bc_u_ND} \\
	&\left.\pad{w}{r}\right|_{r=0}=0, \quad \left.w\right|_{r=1}=-\Kn C\left.\pad{w}{r}\right|_{r=1},\label{bc_w_ND}\\
	&\left.T\right|_{x=0}=1, \qquad \left.T\right|_{x=1}=0.\label{bc_T_ND}
\end{align}
\end{subequations}

\section{Calculation and interpretation of the effective thermal conductivity}
The ETC is defined as the ratio of the heat flux per unit area to the temperature gradient driving
this flux. In non-dimensional form, the ETC may be expressed as
\begin{equation}\label{def_k_eff}
	k_\text{eff}= Q\left(-\nd{T}{x}\right)^{-1},
\end{equation}
where the heat flux per unit area is
\begin{equation}\label{def_Q}
	Q=2\int_0^1 w(r,x)r\, \ud r \, .
\end{equation}
Calculating the integral in \eqref{def_Q} requires solving Eqn.~\eqref{eq_w_ND} for the axial
component of the flux $w$, which is trivial because the temperature does not depend on the radial
coordinate. Upon solving \eqref{eq_w_ND} and applying the boundary conditions
\eqref{bc_w_ND}, we find that
\begin{equation}\label{sol_w_dTdx}
	w(r,x) = -k\left(1-\frac{I_0(r/\Kn )}{I_0(1/\Kn )+C I_1(1/\Kn )}\right)\nd{T}{x},
\end{equation}
where $I_\nu$ is the modified Bessel function of the first kind of order $\nu$. The heat flux per
unit area may then be calculated as in \eqref{def_Q}, which yields
\begin{equation}\label{expr_Q}
	Q=-k\left(1-\frac{2\Kn I_1(1/\Kn )}{I_0(1/\Kn )+C I_1(1/\Kn )}\right)\nd{T}{x}.
\end{equation}
According to its definition in \eqref{def_k_eff}, the ETC is finally given by
\begin{equation}\label{k_nw}
  \frac{k_\text{eff}^*}{k^*}=1-\frac{2\Kn I_1(1/\Kn )}{I_0(1/\Kn )+C I_1(1/\Kn )}
\end{equation}

In principle, the temperature can be calculated from \eqref{expr_Q} by first noting that, at steady
state, the flux $Q$ must be uniform in the axial direction: if a different amount of energy enters
the wire to that leaving, then the temperature must vary with time. Due to the temperature dependence of
the parameters appearing in \eqref{expr_Q}, we have that $T$ satisfies a nonlinear ODE of the form
\begin{equation}\label{ODE_T}
	\nd{T}{x}=f(T,Q),
\end{equation}
which is analogous to the Reynolds equation in fluid mechanics \cite{Ockendon1995}. Solving the
first-order ODE \eqref{ODE_T} and imposing the boundary condition $T = 1$ at $x = 0$ determines the
temperature in terms of the flux $Q$. Subsequently applying the boundary condition $T = 0$ at $x =
1$ to the solution for the temperature enables the flux to be obtained.

Due to the dependence of $\Kn=\ell^*(T^*)/R^*$ on temperature and the radius of the NW, Eqn.~\eqref{k_nw} can be
reduced to simpler expressions in some limiting cases. For instance, at very low temperatures or for
very small radii we have $\Kn\gg1$. Using $I_0(\xi)=1+O(\xi^2)$ and $I_1(\xi)=\xi / 2 +O(\xi^3)$ for
$\xi\ll1$, as well as $C(\Kn) = 1 - \Kn^{-1} + O(\Kn^{-2})$ for $\Kn \gg 1$, we find that
\eqref{k_nw} can then be reduced to
\begin{equation}\label{k_eff_Kn_large}
  \frac{k_\text{eff}^*}{k^*}=\frac{1}{2\Kn}+O\left(\frac{1}{\Kn^2}\right), \quad \Kn \gg 1,
\end{equation}
from which we deduce $k_\text{eff}^*\propto R^*$, in agreement with previous theoretical results
\cite{Ma2012,Alvarez2008,Tzou2011,Alvarez2009}. In deriving the leading-order term in
\eqref{k_eff_Kn_large}, only the leading contribution to $C$, given by $C \sim 1$, is
required. Thus, any alternative expression of $C$ that has a large-$\Kn$ expansion of the form $C =
1 + O(\Kn^{-\alpha})$, with $\alpha > 0$, will result in the asymptotic behaviour
$k_\text{eff}^*/k^* \sim (2\Kn)^{-1}$ as $\Kn \to \infty$. For larger devices or at high
temperatures, where $\Kn \ll 1$, the ETC should reduce to its classical value as non-local effects
become negligible. To show this, we use the relation \cite{Segura2011} $I_{1}(\xi) / I_0(\xi) = 1 -
(2\xi)^{-1} + O(\xi^{-2})$ for $\xi \gg 1$ in \eqref{k_nw} to obtain
\begin{equation}
	\frac{k_\text{eff}^*}{k^*}= 1-2\Kn+O\left(\Kn^2\right), \quad \Kn \ll 1;
        \label{k_eff_Kn_small}
\end{equation}
hence the conductivity tends to the bulk value as $\Kn \to 0$.
The asymptotic behaviour $k_\text{eff}^*/k^* \sim 1-2\Kn$ as $\Kn \to 0$ will be true for all
functional forms of $C$ with the limit $C \to 0$ as $\Kn \to 0$.

\begin{figure}
  \centering
  \begin{subfigure}{0.49\textwidth}
    \centering
    \includegraphics[width=\textwidth]{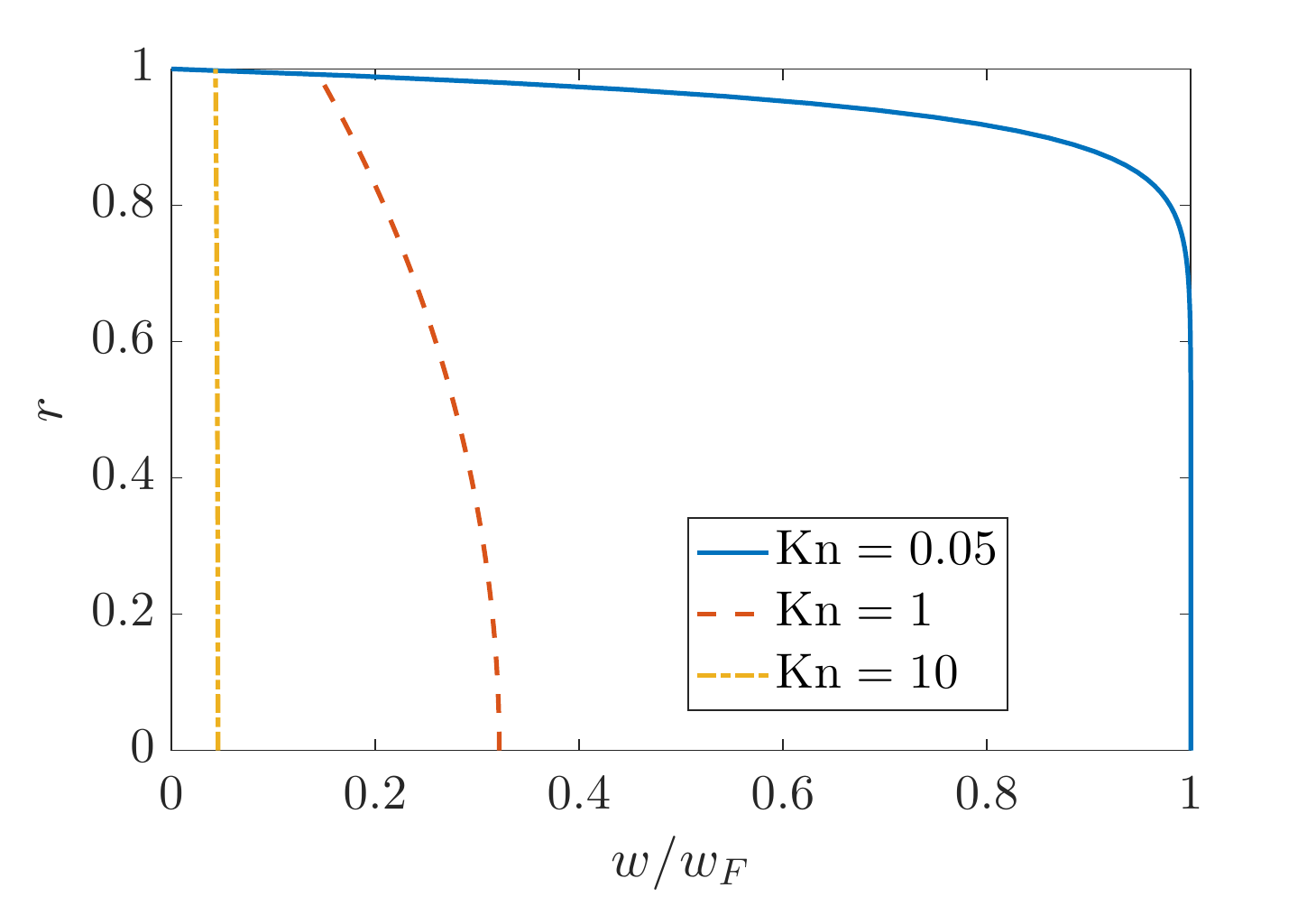}
    \caption{}
  \end{subfigure}
  \begin{subfigure}{0.49\textwidth}
    \centering
    \includegraphics[width=\textwidth]{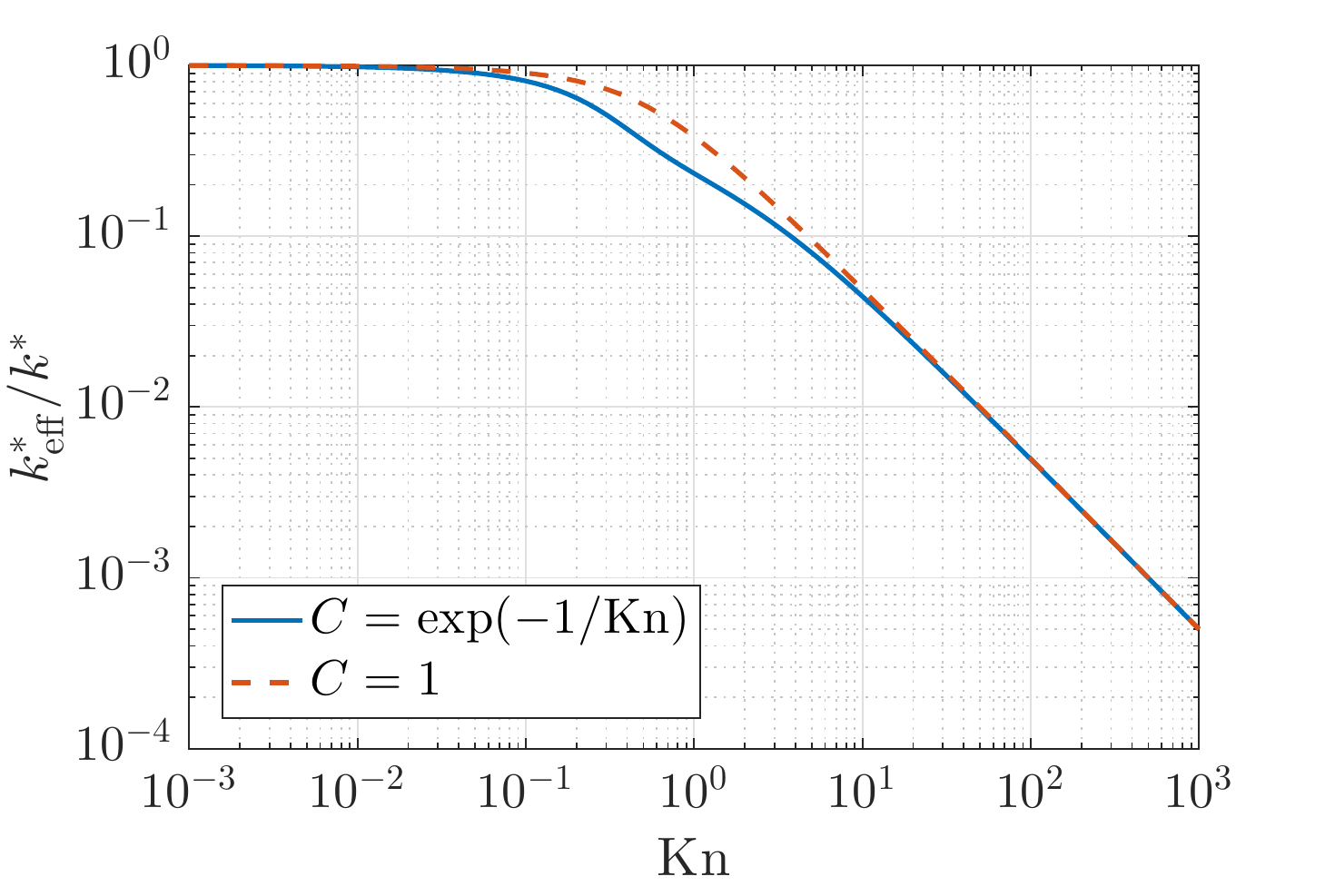}
    \caption{}
  \end{subfigure}
  \caption{(a) Axial component of the heat flux according to Eqn.~\eqref{sol_w_dTdx} with $C = \exp(-1/\Kn)$ for different values of $\Kn$ and scaled with the classical heat flux $w_\text{F}$. Solid, dashed and dashed-dotted lines correspond to $\Kn=0.05$, $\Kn=1$ and $\Kn=10$ respectively. (b) The dependence of the ETC given by \eqref{k_nw} on the Knudsen number for two choices of the function $C$.}
  \label{fig:w_k}
\end{figure}

To aid with the physical interpretation of the model solutions and their asymptotic limits, the
radial dependence of the axial flux $w$ is plotted for three Knudsen numbers 
($\Kn = 0.05$, $1$, and $10$) in Fig.~\ref{fig:w_k} (a). 
The ETC, given by \eqref{k_nw}, is plotted as a function of the Knudsen number in
Fig.~\ref{fig:w_k} (b) for two choices of the function $C$. 
The figures make it clear that there are three distinct regimes to
consider, corresponding to diffusive ($\Kn \ll 1$), ballistic ($\Kn \gg 1$), and mixed ($\Kn =
O(1)$) modes of thermal energy transport. 
The three values of the Knudsen number in
Fig.~\ref{fig:w_k} (a) are chosen to clearly illustrate how the axial flux varies across these three
regimes.

For small Knudsen numbers, $\Kn \ll 1$, the phonon MFP is
small compared to the radius of the NW. Phonons in the bulk are therefore more likely to collide
with each other before reaching the boundary of the NW and scattering. This corresponds to the
boundary condition $w\to0$ as $\Kn\to0$. Due to the relatively small influence of non-local effects,
thermal energy in the bulk is transported with little resistance across the NW, with an axial flux that
approximates the classical flux $w_F$ predicted by Fourier's law. However, there is a thin boundary layer near the edge of
the NW where the flux rapidly decreases to zero in order to satisfy the no-slip condition. The
(dimensional) width of this boundary layer is $O(\ell^*)$, reflecting the fact that boundary effects
become relevant on length scales that are commensurate with the phonon MFP. As the Knudsen number
tends to zero, the width of this boundary layer does so as well. The restricted transport of thermal
energy in the boundary layer leads to a slight reduction in the ETC compared to its bulk value,
which is captured in the small-$\Kn$ limit of $k_\text{eff}$ given by
\eqref{k_eff_Kn_small} and shown in Fig. \ref{fig:w_k} (b) by the slow reduction in
$k_\text{eff}^*/k^*$ as Kn increases from $10^{-3}$ to $10^{-1}$.

In the limit of large Knudsen number, $\Kn \gg 1$, the phonon MFP greatly exceeds the radius of the
NW. Phonons therefore collide more frequently with the boundary than with each other, resulting in a
substantial decrease in the flow of thermal energy. The strong influence of non-local effects and
large slip length lead to an axial flux that is uniform along the radial direction and, in
particular, close to its value on the boundary, $w \sim \Kn^{-1} w_F / 2$. Therefore, as $\Kn \to
\infty$, corresponding to an arbitrarily large MFP relative to the NW radius, the axial flux tends
to zero. The restricted transport of energy in this regime results in a marked decrease in the ETC,
the magnitude of which also scales linearly with $\Kn^{-1}$, as demonstrated in
Eqn.~\eqref{k_eff_Kn_large}.

The proportionality between the flux $w$ and the inverse Knudsen number $\Kn^{-1}$ can be understood
by drawing on the analogy between nanoscale heat transfer and fluid dynamics. In the limit of large
$\Kn$ (analogous to large viscosity), the net force exerted by the temperature gradient (analogous
to pressure) on an arbitrary cross section of the NW must balance the net traction (i.e., shear
stress) along its circumference,
\begin{align}
\pi k \nd{T}{x} \sim 2 \pi \Kn^{2} \left.\pad{w}{r}\right|_{r = 1}.
  \label{stress_bal}
\end{align}
From Eqn.~\eqref{stress_bal} it is seen that the rate of shear strain, $\partial w / \partial r$,
induced by the temperature gradient is small and of size $O(\Kn^{-2})$. The scaling for the flux
(analogous to velocity) is determined by the slip condition \eqref{bc_w_ND}, yielding $w \sim
\Kn\, \partial w / \partial r \sim \Kn^{-1} w_F / 2$, where the rate of strain has been eliminated
using \eqref{stress_bal} and $C$ has been approximated by its leading-order contribution $C \sim 1$
as $\Kn \gg 1$. Thus, the scaling with $\Kn^{-1}$ arises from the combination of large $O(\Kn)$ slip
length and small $O(\Kn^{-2})$ rate of strain.


The case of $\Kn = O(1)$ corresponds to an intermediate regime where the length scales of the NW
radius and MFP are the same order of magnitude. Boundary scattering plays an important role and the
order-one slip length leads to a parabolic-like flux. Unlike the large- and small-$\Kn$ regimes, the
ETC is strongly dependent on the functional form of $C$ when $\Kn = O(1)$. Figure \ref{fig:w_k} (b)
shows the ETC when $C = \exp(-1 / \Kn)$ (solid line) and $C = 1$ (dashed line). The former
corresponds to the situation where the slip length $\ell_s^*$ is smaller than the non-local length
$\ell^*$ (recall that $\ell_s^* = \ell^* C$). Taking $C = 1$ implies that the slip length and MFP
are equal. The exponential form of $C$ results in a slight kink near $\Kn = 1$. In physical terms,
this kink is associated with a reduced ETC compared to the $C = 1$ case and is the consequence of a
smaller slip length and hence axial flux. Using $C = 1$ smooths out this kink and yields higher
values of the ETC due to greater slip.

\section{Model validation and comparison}
\label{sec:valid}

\begin{figure}
	\centering
	\begin{subfigure}{0.48\textwidth}
		\centering
		\includegraphics[width=\textwidth]{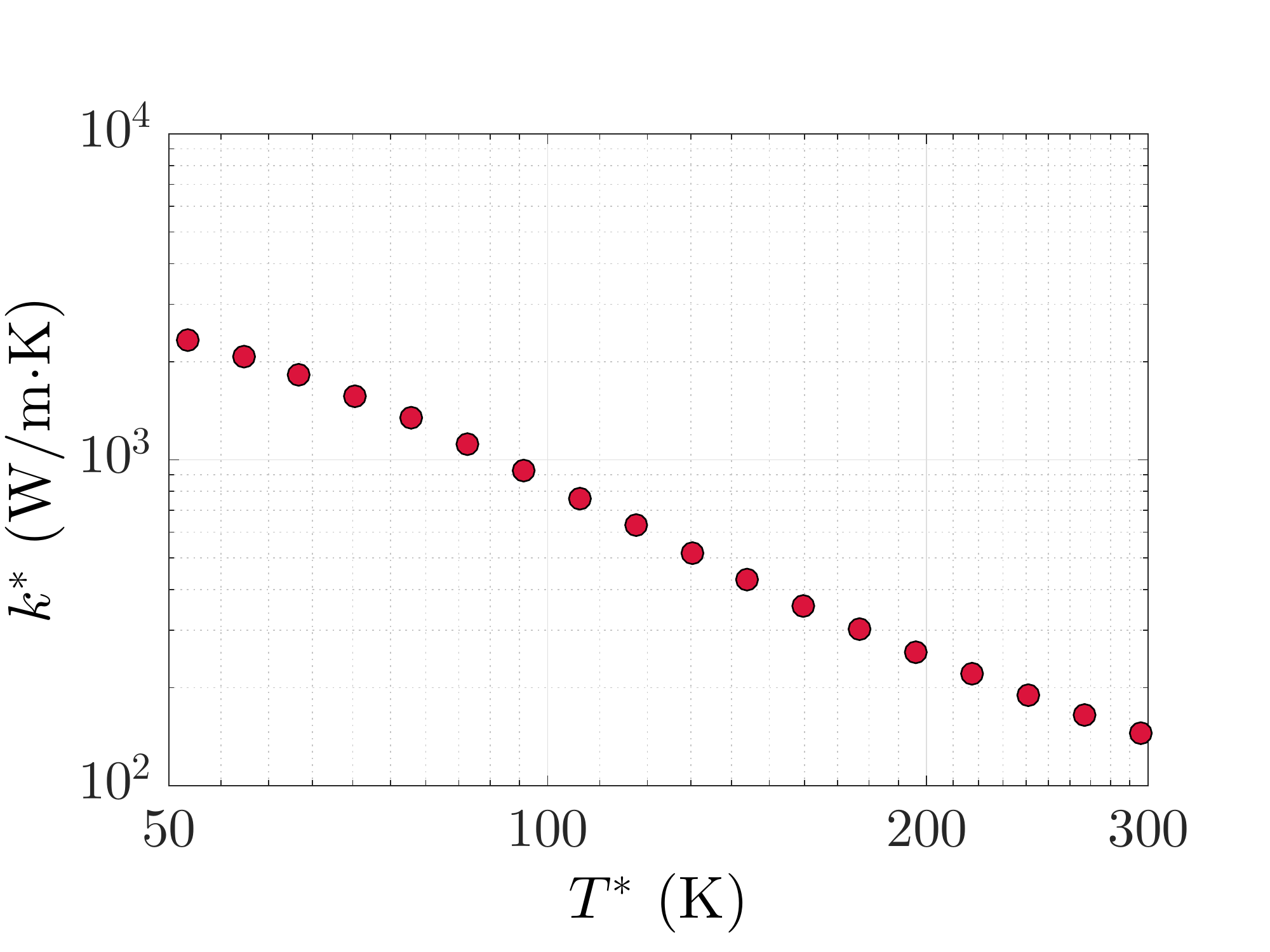}
		\caption{}
	\end{subfigure}
	~
	\begin{subfigure}{0.48\textwidth}
		\centering
		\includegraphics[width=\textwidth]{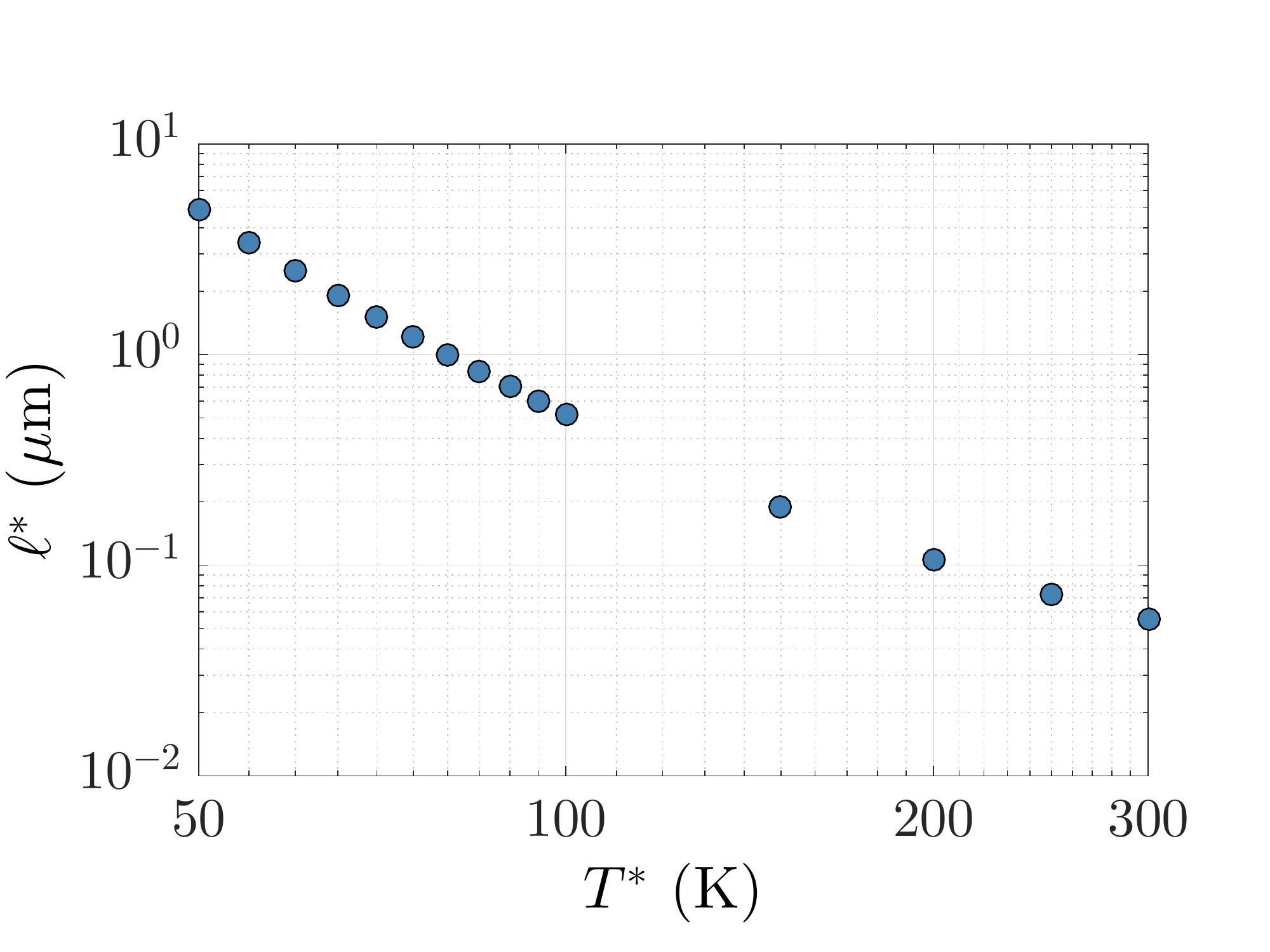}
		\caption{}
	\end{subfigure}
	\caption{Temperature dependence of (a) the bulk thermal conductivity and (b) the non-local length for Si.}
	\label{fig:k_ell}
\end{figure}

The prediction of the ETC is now compared against experimental data measured from Si nanowires with
diameters $D^* = 2R^* = 37$ nm, $56$ nm, and $115$ nm at temperatures ranging from $T^* = 50$ K up
to $300$ K \cite{Li2003}. Evaluating the ETC prediction requires knowledge of the temperature
dependence of the non-local length $\ell^*$ and thermal conductivity $k^*$. These two quantities are
obtained from \textit{first principles} in the KCM framework~\cite{Torres2016} using an open-source
code~\cite{Torres2017} and plotted as functions of temperature in Fig.~\ref{fig:k_ell}. The bulk MFP
ranges from 55 nm to 5 $\mu$m, corresponding to Knudsen numbers between 2.98 and 262 ($D^* = 37$
nm), 1.97 and 173 ($D^* = 56$ nm), and 0.958 and 84.2 ($D^* = 115$ nm). Thus, the exponential form
of $C$ will be relevant when comparing the ETC prediction to the experimental data. Indeed, we find
poor agreement when the the dependence of $C$ on $\Kn$ is neglected and $C$ is held at unity.

\begin{figure}
\begin{subfigure}{0.49\textwidth}
	\includegraphics[width=\textwidth]{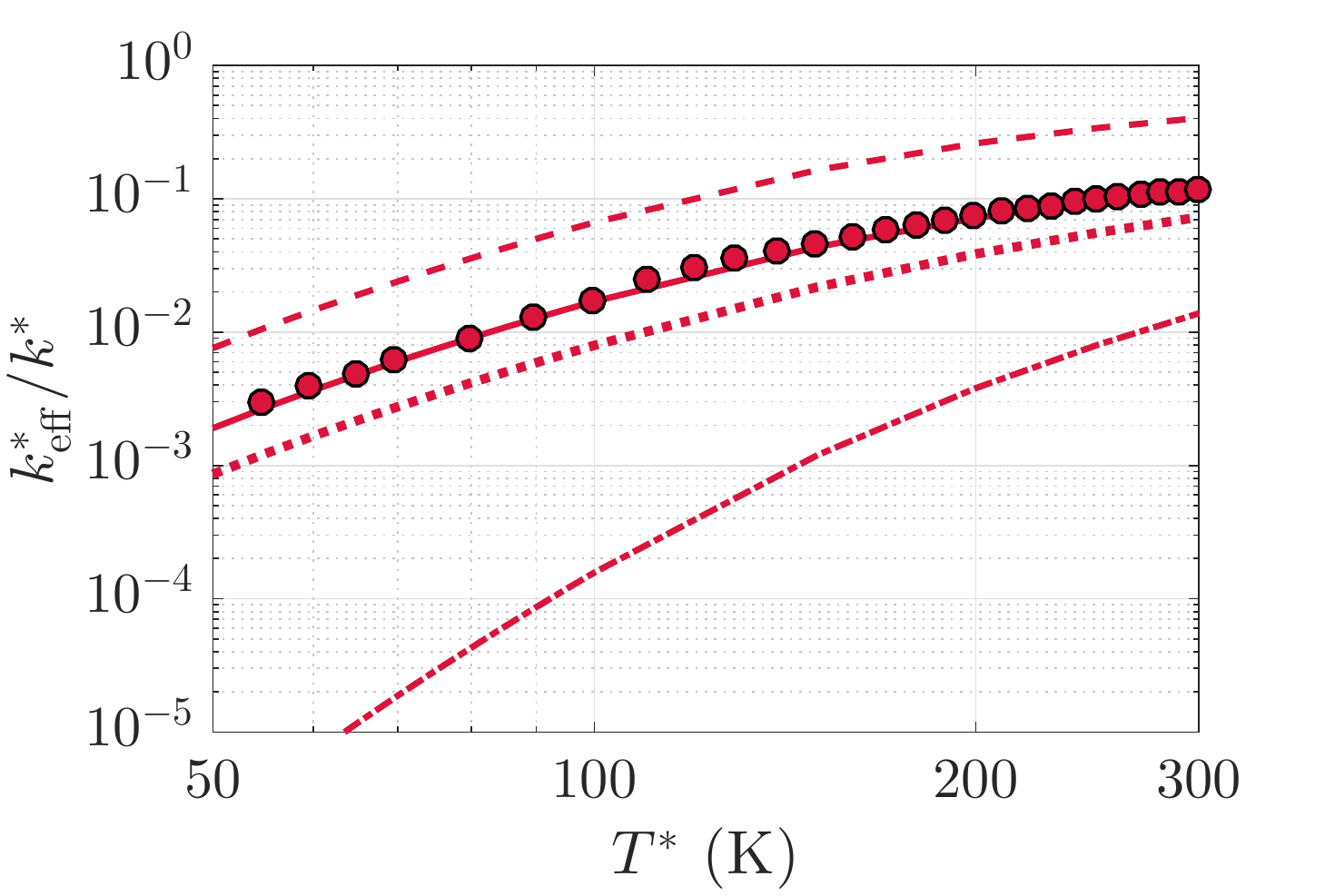}
	\caption{$D^*=37$ nm.}
\end{subfigure}
\begin{subfigure}{0.49\textwidth}
	\includegraphics[width=\textwidth]{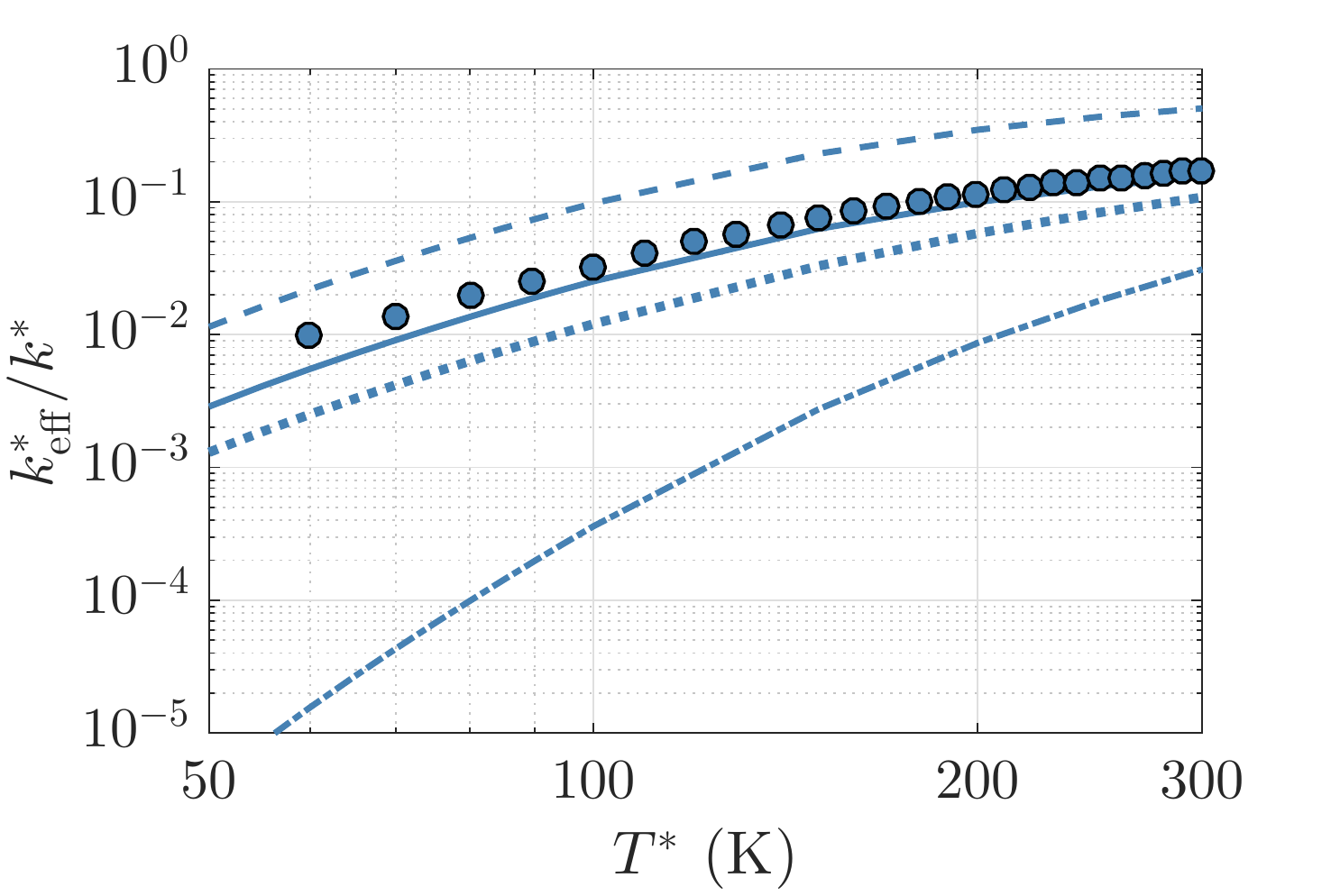}
	\caption{$D^*=56$ nm.}
\end{subfigure}
\begin{subfigure}{0.49\textwidth}
	\includegraphics[width=\textwidth]{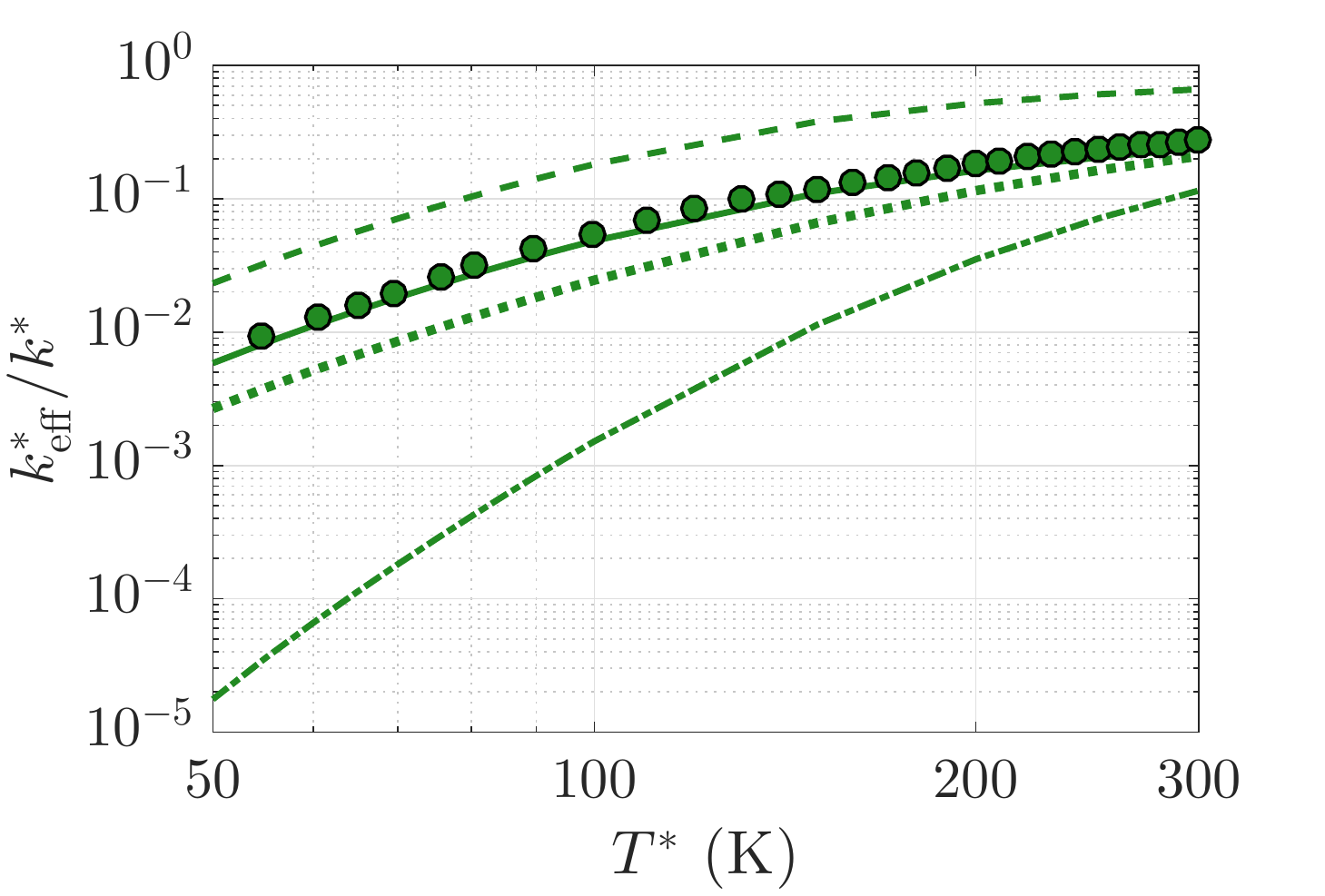}
	\caption{$D^*=115$ nm.}
\end{subfigure}
\hspace{0.2cm}
\begin{subfigure}{0.49\textwidth}
	\includegraphics[width=\textwidth]{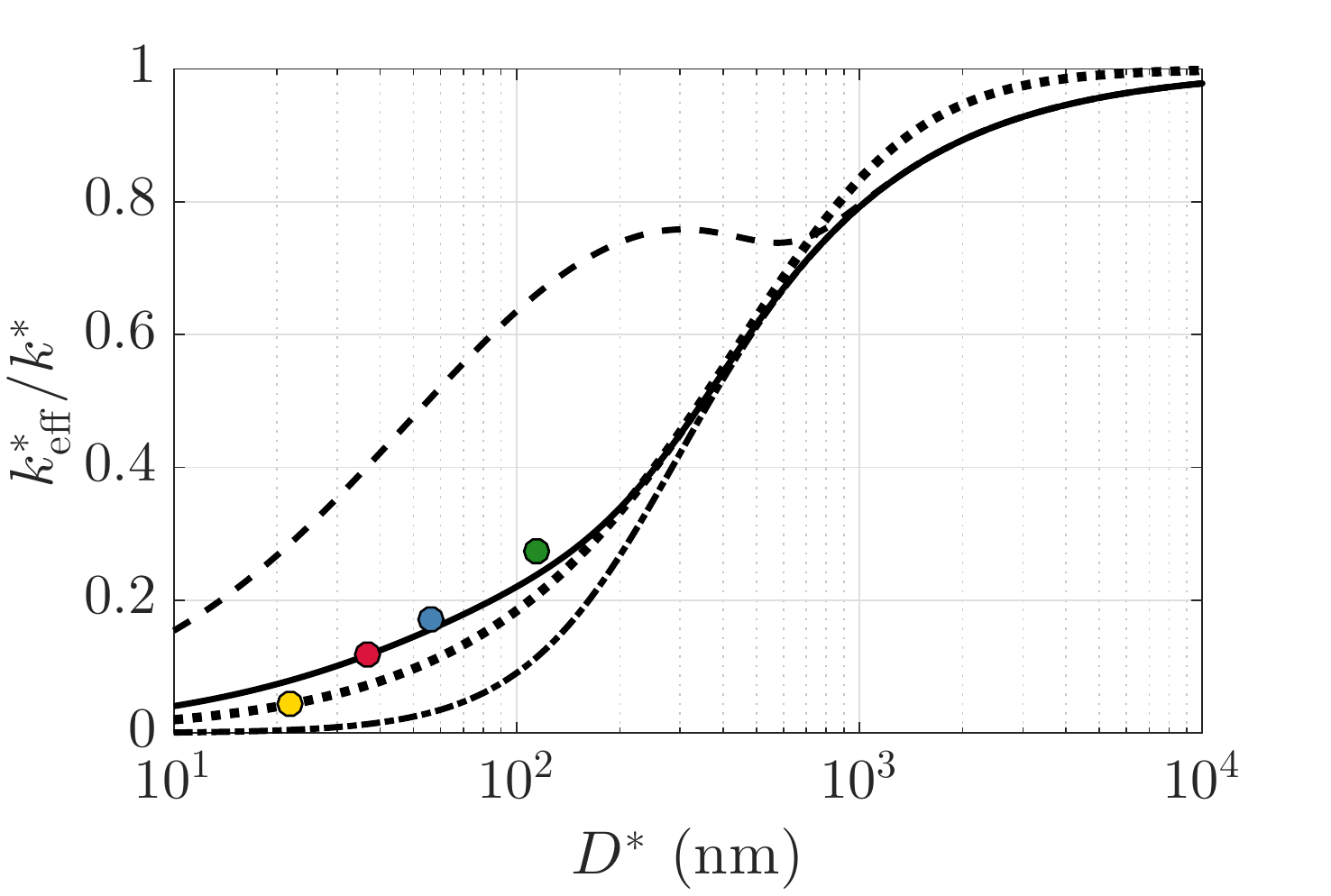}
	\caption{$T^*= 300$ K.}
\end{subfigure}
\caption{Comparison of theoretical predictions and experimental data for $k_\text{eff}^*$ in Si NWs of varying radii and temperature. Solid lines correspond to Eqn.~\eqref{k_nw}; dashed, dotted and dashed-dotted lines represent the model predictions given by Refs.~\cite{Ma2012,Alvarez2008,Dong2014}, respectively; and circles denote experimental data \cite{Li2003}.}
  \label{fig:Comp_k_nw}
\end{figure}

Figure \ref{fig:Comp_k_nw} shows the ETC predicted by \eqref{k_nw} (solid lines) and experimentally
measured values of the ETC (symbols) as functions of temperature for three NW diameters
(Figs.~\ref{fig:Comp_k_nw} (a)--(c)) and as functions of the diameter for a fixed temperature
(Fig.~\ref{fig:Comp_k_nw} (d)). The ETC $k_\text{eff}^*$ is normalised against the bulk value $k^*$
to highlight the relative change that occurs as the diameter of the NW and non-local length
decrease. In all cases, the theoretical predictions are in excellent agreement with the experimental
data.

The dashed, dotted, and dashed-dotted lines in Fig.~\ref{fig:Comp_k_nw} represent theoretical
predictions of the ETC computed from existing models found in
Refs.~\cite{Ma2012,Alvarez2008,Dong2014}, respectively. It is clear that that the model proposed
here consistently yields predictions that are more representative of the experimental data. The
models of Ma \cite{Ma2012} and Dong \emph{et al.}~\cite{Dong2014}, both of which are based on the
G-K equation, are unable to quantitatively capture the experimental data associated with all three
NW diameters.  These inaccuracies can be attributed to the different boundary conditions that are
used in their models. Dong \emph{et al.}~\cite{Dong2014} employ a no-slip boundary condition,
equivalent to $C=0$ in our model. Imposing this boundary condition leads to particularly large
errors at low temperatures (or large Knudsen numbers). Moreover, the calculated ETC becomes
quadratically dependent on the NW radius, i.e., $k_\text{eff}^* / k^* \propto R^{*2}$, which
contradicts the experimentally measured linear dependence between these two quantities. Ma
\cite{Ma2012} instead considers a prescribed flux on the boundary, which leads to substantial
inaccuracies across the whole temperature range. 
The expression for the ETC presented by Alvarez and Jou \cite{Alvarez2008} gives the most accurate
prediction of the three existing models, although it generally underestimates the experimentally
measured values. A distinguishing feature of Alvarez and Jou's expression is that it is derived from
the linearised Boltzmann equation rather than the G-K equation.  This derivation does not explicitly
account for the cylindrical geometry of the nanowire and thus no boundary conditions can be imposed
along the outer edge, which may contribute to the errors seen in Fig.~\ref{fig:Comp_k_nw}.  
In comparison to the models derived from the G-K equation, the model of Alvarez and Jou predicts
that the ETC converges more rapidly to the macroscopic value with increasing NW diameter; 
see Fig.~\ref{fig:Comp_k_nw} (d). However, the lack of experimental data for microscale NW 
diameters  prevents the models from being validated in this regime.

\section{Conclusion}

We have proposed a new phenomenological slip model that can be coupled with the Guyer--Krumhansl
equation in order to obtain an analytical expression for the size-dependent ETC in thin nanowires. A
key advantage of the proposed model is that it only requires knowledge of the bulk thermal
conductivity and MFP and does not involve any fitting parameters. The prediction of the ETC of Si
nanowires is found to be in excellent agreement with experimental measurements, which is
particularly remarkable given the simple nature of the underlying model. When compared against
existing models for the ETC, we observed that the proposed model yielded the most accurate
predictions in all cases.

Crucial to the success of our model is the condition imposed on axial flux $w$ at the outer boundary
of the nanowire. Similar models, but with different boundary conditions, were shown to be in poor
agreement with much of the experimental data. It is surprising that the present choice, based purely
on an analogy with viscous fluid flow, works so well without resorting to any parameter
fitting. Consequently, in our future work, we intend to further investigate the precise form of the
boundary condition and, ideally, produce an expression based on a detailed physical model.

Finally, although we focused on circular nanowires, the model could, in principle, be applied to
arbitrary geometries. Analytical solutions could be sought in the case of simple geometries such as
rectangular nanofilms. In the case of complex geometry, the model could be straightforwardly
implemented in finite element software, enabling a broad range of scenarios to be explored.




\bibliographystyle{unsrtnat}
\bibliography{ETC.bib}

\end{document}